\documentstyle[11pt,oldlfont]{article}
\def\Re{\rlap{\rm I}\mkern3mu{\rm R}}
\def\Zed{\rlap{\rm \mbox{\sf Z}}\mkern3mu{\rm \mbox{\sf Z} }}

\def\db{\mbox{\boldmath$\delta$}}

\def\ex{\mbox{\boldmath$d$}}

\def\bi{\bibitem}

\begin{document}

\thispagestyle{empty}

\begin{flushright} hep-th/0204179 \\
                   AEI-2002-033  \end{flushright}

\vspace*{0.2cm}

\begin{center}{\LARGE {Black hole entropy and thermodynamics from
symmetries}}   

\vskip0.2cm

Sebasti{\'a}n Silva

\vskip0.2cm

Max Planck Institut f{\"u}r Gravitationsphysik, Albert Einstein Institut,\\
Am M{\"u}hlenberg 5, D-14476 Golm, Germany

\vskip0.2cm

\begin{minipage}{12cm}\footnotesize

{\bf ABSTRACT}
\bigskip

Given a boundary of spacetime preserved by a 
{\it Diff}$(S^{1})$ sub-algebra, we propose  a systematic method to compute
the zero mode and the central extension of the associated Virasoro
algebra of charges.
Using these values in the Cardy formula, we may derive an associated
statistical entropy to be compared with the Bekenstein-Hawking result.

To illustrate our method,
we study in detail the BTZ and the rotating
Kerr-adS$_{4}$ black holes (at spatial infinity and on the 
horizon). In both cases, we are able to reproduce the area law with the
correct factor of $1/4$ for the entropy. We also recover within our 
framework the first law of black hole thermodynamics.

We compare our results with the analogous derivations proposed by
Carlip and others. Although similar, our method differs in the
computation of the zero mode. In particular, the normalization
of the ground state is automatically fixed by our construction.

\bigskip
\end{minipage}
\end{center}
\newpage

\section{Introduction}\label{intro}

There have been a lot of interest in understanding black hole entropy
from a microscopical description, using either string theory \cite{SV}
or loop quantum gravity \cite{ABCK}.  A third symmetry-based approach
originated by  
Strominger \cite{Str} and Birmingham, Sachs and Sen \cite{Birmingham:1998jt} is
attracting more and more interest \cite{Carlip:1998qw}-\cite{Carlip:2002be}. 
This
method assumes that the symmetries of a black hole
horizon are good enough, first to survive a
quantification, and second to count the density of states at a given energy.
The main argument in favor of these new ideas is probably that they 
seem to work, although we still do not clearly understand why.

Let us start by summarizing the key points of this calculation: 
Given a boundary of spacetime
(spatial infinity for Strominger \cite{Str} or the black hole horizon in Carlip
 \cite{Carlip:1998wz,Carlip:1999cy} and Solodukhin
\cite{Solodukhin:1998tc} works), we need to identify the  
diffeomorphisms which preserve the boundary 
conditions. Let us assume that their associated algebra contains at least one
sub-algebra isomorphic to  {\it Diff}$(S^{1})$. This sub-algebra can be
parametrized 
by a (infinite) discrete set of vector fields $\xi ^{\mu } _{n} (x)$, 
$n\in \Zed$ that satisfy 
\begin{equation}\label{diff1}
i \left\{ \xi _{m}, \xi _{n} \right\} =  (m-n)\xi _{m+n},
\end{equation}
where 
$\left\{ \ , \  \right\}$ denotes the Lie bracket between
vector fields (the spacetime index $^{\mu }$ has been omitted).

The next step consists in constructing, either using
an Hamiltonian \cite{RT,BHT} or a covariant-Lagrangian formalism
\cite{KBL}-\cite{Barnich:2001jy}, the charges 
associated with each of these vector fields:
\begin{equation}\label{chargvect}
\xi _{n} (x) \leftrightarrow Q_{n}.
\end{equation}
This is usually a difficult and ``boundary condition dependent''
task. Indeed, only in the cases where the asymptotic conditions
which generate the algebra (\ref{diff1}) are
fully understood, $Q_{n}$ can be derived methodically.

But let us assume that we found these charges. 
A bracket is then defined by $\left[Q_{m}, Q_{n}
\right]:= \delta _{\xi_{m}} Q_{n}$ where $\delta _{\xi_{m}}$ is the
Lie derivative along $\xi_{m}^{\mu}$ which acts only on the
fields (but not on the vectors $\xi ^{\mu }_{n}$). In this situation, a
notable theorem by Brown and 
Henneaux \cite{BHT} shows an isomorphism
between the two brackets $\{. , .  \}$ and $[. , . ]$ up to central
charges (a Lagrangian version of this theorem is derived in
\cite{SiT,Barnich:2001jy}). In particular, from the algebra
(\ref{diff1}), we get,  
\begin{equation}\label{isomorphic}
i \left[ Q_{m},Q_{n} \right] =  (m-n) Q_{m+n}
+  \frac{c}{12} m (m^{2} -1) \delta _{m+n},
\end{equation}
for some central extension $c$ to be determined. 
We emphasize here that the freedom
of shifting $Q_{0}$ by a constant has been fixed by the
contribution linear in $m$ to the central charge (and then $Q_{-1}$,
$Q_{0}$ and $Q_{1}$ form a proper $sl (2,\Re)$ sub-algebra). 

Once  $Q_{0}$ and $c$ are found, the next idea is to insert them in a
Cardy-like formula \cite{Cardy:ie} to find the entropy of the black hole
\begin{equation}\label{cardy}
S = 2 \pi \sqrt{\frac{c}{6} Q_{0}}.
\end{equation}
The last step is simply to
check whether 
this calculation reproduces or not the well-known semi-classical result.

It is still under discussion why the formula (\ref{cardy}) would
or should work\footnote{Using either the
derived $Q_{0}$ and $c$, or some ``effective'' ones.} (see the nice references
\cite{Carlip:1998qw,Carlip:1999cy,Carlip:2000nv}). In fact, the
Cardy's result comes from a quantum 
calculation in a two dimensional 
conformal field theory which has not been identified in our
gravitational theory. Moreover, it is not obvious why the {\it
classical} central charge of equation (\ref{isomorphic}) would then be
protected by quantum corrections. Therefore,
only a full understanding of the quantum theory which governs the
black hole horizon would be able to answer
these questions. 

However, following Strominger \cite{Str}, we can still investigate
whether, when and how this Cardy equation works.
This is the purpose of this manuscript.
To proceed, we first propose a straightforward method (based on one
single equation) to compute $Q_{0}$ and
$c$ on the horizon of any $D$-dimensional black hole ($D\geq 3$). We
then  show in 
specific examples that these derived $Q_{0}$ and $c$ indeed reproduce
the area law formula (with the correct normalization) when used in
equation (\ref{cardy}). Our construction is strongly
inspired from Carlip's results 
\cite{Carlip:1998wz,Carlip:1999cy}, although it differs in the
construction of $Q_{0}$. A detailed 
comparison with these works is carried out.

In section \ref{shortcut}, we explain and motivate our new
construction of  $Q_{0}$ and $c$, and compare it with previous results
found in the literature (see also appendix A). In section \ref{exa},
we study the examples of BTZ black hole at spatial infinity and on
the horizon (the Carlip boundary conditions are discussed in appendix B and
the Eddington-Finkelstein change of coordinates in appendix C). 
In both cases, we find that equation (\ref{cardy}) correctly reproduces
the Bekenstein-Hawking entropy. We then show how the first law of black
hole thermodynamics is embedded in our framework.
Finally, we extend our investigation to a general four-dimensional Kerr-adS
black hole (which covers Schwarzschild, Kerr and adS-Schwarzschild
black holes). We again find a perfect agreement for the entropy
computed on the horizon using the Cardy formula.

\section{A direct derivation of $Q_{0}$ and $c$}\label{shortcut}

The charges associated with the
diffeomorphism invariance of general relativity depend on the boundary
considered. In concrete, there is {\it no} 
general formula like $Q=\int_{bd} {\cal Q}[g]$, for some density 
 ${\cal Q}[g]$, which would be defined independently of the boundary
``{\it bd}''. 
 This is due to the fact that an unambiguous 
construction of ${\cal Q} [g]$ depends necessarily
on the boundary 
conditions\footnote{A general formula for the charges was recently given in
\cite{Barnich:2001jy} following the methods of Anderson and Torres \cite{AT}. 
This formalism requires that the metric approaches a given background
metric on the boundary. 
However in some cases as Brane-World scenarios,
no background metric is needed in order to define conserved charges 
\cite{Silva:2000ys}.} \cite{RT,BHT}. 

On the other hand, a general boundary-independent formula for the {\it
variation} of the charge $\db {\cal Q} [g]$ can be derived from first
principles \cite{RT,ABR,Wa,Jezierski:vu,Si}. This formula depends {\it only on the gauge symmetry and on
the equations of motion} of the theory \cite{Si}.
For 
gravity in any spacetime dimension $D \geq 3$ with or without
cosmological constant, the
variation of the charge $Q_{n}$ associated with the diffeomorphism
$\xi ^{\mu }_{n} (x)$ is given by \cite{JS2,Silva:2000ys}:
\begin{equation}\label{algebra}
\db  Q_{n } = 
\int_{\it bd} \frac{1}{16 \pi G} \left( 2\  \nabla_{\sigma }
\xi^{\tau}_{n} 
\db \left( \sqrt{\left| 
g\right|} g^{\sigma [\mu }  \delta^{\nu] }_{\tau
} \right) + 6\  \db  \left( \Gamma^{\tau }_{\rho \sigma}\right)
\sqrt{\left| 
g\right|} g^{\sigma [\mu}  \delta^{\nu}_{\tau } \xi_{n}
^{\rho ]}  \right) dS_{\mu \nu }
\end{equation}
with $dS_{\mu \nu }$ the bi-normal to the intersection
between the spacetime boundary considered and a (partial) Cauchy
hypersurface.

The equation (\ref{algebra})  is valid on  {\it any boundary}
of spacetime whose boundary conditions are compatible with the
variational principle\footnote{A given set of boundary conditions are
compatible with the variational principle if 
there exits a Lagrangian ${\cal L}$ in the class ${\cal L} \sim {\cal L} +
\partial_{\mu } \Sigma ^{\mu}$, such that on-shell $\delta \int 
{\cal L} = 0$ (with no boundary terms).}. In particular, it has been used at 
spatial infinity \cite{JS2} to recover a covariant formulation of the ADM
mass \cite{KBL} and also on a Brane-World 
to properly define its associated conserved charges \cite{Silva:2000ys}. 
The purpose of this paper is to study this equation
(\ref{algebra}) on the horizon of a 
black hole. We then assume that there exits a properly defined set of boundary
conditions on the horizon compatible with the variational principle
(although we will not need to find it explicitely).
The main point is that equation (\ref{algebra})  will then allow us first, to
compute directly the zero 
mode $Q_{0}$ and the central charge $c$ and second,
to derive the first law of black hole thermodynamics.

To proceed, let us 
use the general formula (\ref{algebra}) 
with $\db = \delta _{\xi_{m}}$ that is a
Lie derivative along $\xi_{m}^{\mu } (x)$ given by
\begin{equation}\label{known}
\delta_{\xi} g_{\mu \nu } = 2 \nabla_{( \mu} \xi _{\nu)  } ,\ \   \delta_{\xi} \Gamma^{\tau }_{\rho \sigma} =
\frac{1}{2}g^{\tau \nu} 
\left(\nabla_{\rho} \delta _{\xi} g_{\sigma \nu} + \nabla_{\sigma}
\delta _{\xi} g_{\rho \nu} - \nabla_{\nu} \delta _{\xi} g_{\rho \sigma
}\right).
\end{equation}
Dropping the following total derivative
\begin{equation}\label{totder}
 \frac{6}{16 \pi G} \partial_{\rho} \left( \sqrt{\left| 
 {g}\right|} \xi_{n} ^{[ \mu} 
{\nabla}^{\nu 
} \xi_{m} ^{\rho ] }\right),
\end{equation}
the  equation (\ref{algebra}) then reduces to:
\begin{eqnarray}
\left[ Q_{m },Q_{n }\right] &=& \delta _{\xi_{m}} Q_{n} \nonumber \\
&=& \int_{\it bd} \frac{\sqrt{\left|  {g}
\right|}}{16 \pi G} \left[ {\nabla} ^{\mu } \xi_{n} ^{\nu }
 {\nabla}_{\rho } \xi_{m} ^{\rho} -  {\nabla} ^{\mu } \xi_{m} ^{\nu }
 {\nabla}_{\rho } \xi_{n} ^{\rho} + 2  {\nabla} _{\rho} \xi_{n} ^{\mu}
 {\nabla}^{\rho} \xi_{m} ^{\nu} \right. \nonumber\\
 & & \ \ \ \ \ \  \left. + ( 
{\cal R}^{\mu \nu}{}_{ \rho \sigma } - 2 {\delta }^{\mu}_{\rho} 
{{\cal R}} ^{\nu}{}_{ \sigma}) \xi_{n} ^{\rho }\xi_{m}
^{\sigma }  - (\mu \leftrightarrow \nu) \right] dS_{\mu \nu }, \label{findth}
\end{eqnarray}
with ${\cal R}^{\mu \nu}{}_{ \rho \sigma }$ the Riemann tensor of the
metric.

Results similar 
to equation (\ref{findth}) can be found 
in the literature
\cite{Koga:2001vq,Barnich:2001jy}, and are described in the appendix A. 
In addition, 
an alternative definition 
for the bracket between two charges was given in the reference
\cite{Carlip:1999cy} (see 
equation (\ref{nbrack})). As shown in section  \ref{exa}, this other bracket
fails to reproduce the known charges at spatial infinity. We will also
discuss its applicability on the horizon.

The main point is that equation (\ref{findth}) can be used for any metric
which satisfies Einstein equations, and not necessarily for a
fixed background metric.
Then, given a black hole solution and a subset of
diffeomorphisms $\xi _{n} (x)$, we can compute the left-hand side of equation
(\ref{isomorphic}) for any $n$ and $m$ using the single formula
(\ref{findth}). In particular, we can consider $m=-n=1$ and $m=-n=2$:
\begin{eqnarray}
i \left[ Q_{1},Q_{-1} \right] &=& 2   Q_{0}. \label{ec1}\\
i \left[ Q_{2},Q_{-2} \right] &=& 4  Q_{0} +  \frac{c}{2} \label{ec2}.
\end{eqnarray}
The left-hand side of these equations being known, it is then easy 
to single out  $Q_{0}$ and $c$.

We would like to insist on the following very important point: the
normalization of $Q_{0}$ is fixed in our proposal by requiring that
the charges $Q_{-1}$, $Q_{0}$ and $Q_{1}$ form a $sl (2,\Re )$
sub-algebra. This means that the right-hand side of equations
(\ref{isomorphic}), 
(\ref{ec1}) and (\ref{ec2}) are completely determined and we cannot
shift $Q_{0}$ by some constant anymore. 
The {\it claim} is then that the $Q_{0}$ and the $c$ computed in that
way on the 
horizon are the effective quantities to be used in the Cardy formula
(\ref{cardy}). We show how this works for specific examples in the
next section. Of course, it would be more gratifying 
to have a general proof of this statement; a proof for
instance based on a quantum conformal theory living on
the horizon. This is however out of the scope of this work.  

The problem of the global normalization of $Q_{0}$, and how to fix it,
is obviously a key feature if we want to compute the entropy {\it {\`a} la
Cardy}. The usual Regge and Teitelboim \cite{RT} or the covariant
symplectic methods
\cite{ABR,Wa,Jezierski:vu,Si} give $Q_{0}$ only up to a constant
shift. At spatial 
infinity, this constant is fixed by requiring that the charge
vanishes on a ``natural background'', as for instance Minkowski
spacetime. On the horizon, the problem is more involved because 
it is hard to define a background 
on which $Q_{0}$ would be set to vanish.
For instance, it is not clear to the
author what would be a natural background for a Kerr black
hole, near to the horizon (Minkowski space does not have an internal
boundary).
 Finally, in the reference \cite{Park:2001zn} the arbitrary shift in
$Q_{0}$ was fixed {\it a posteriori} so that the Cardy formula gave
the expected Bekenstein-Hawking result. However, this approach is
quite unsatisfactory in 
the sense that the non-trivial check of the entropy calculation is lost.

{\it In summary}, given a black hole solution and a set of vector
fields $\xi ^{\mu }_{n} (x)$ satisfying equation (\ref{diff1}), we
can directly compute $Q_{0}$ and $c$ (and then the entropy $S$) by
making use of equations (\ref{findth}), (\ref{ec1}) and (\ref{ec2}).

\section{Examples}\label{exa}

In the following examples we use the method given in the
previous section in order to compute $Q_{0}$ and $c$. 
We start with the rotating BTZ black hole in three
dimensions, at spatial infinity and on the horizon. We then study the four
dimensional adS-Kerr solution. 
We simultaneously compare our results with recent works on the subject 
\cite{Carlip:1998wz,Carlip:1999cy,Dreyer:2001py}.
Finally, we will also use equation
(\ref{algebra}) to derive the (local) first law of black
hole thermodynamics. 

\subsection{The BTZ black hole at spatial infinity}

The metric of the rotating BTZ black hole in Schwarzschild-type
coordinates is given by \cite{Banados:wn}:
\begin{equation}\label{BTZ}
ds^{2} = - N^{2} dt^{2} +  N^{-2} dr^{2} +r^{2} \left(N_{\phi} dt
+d \phi\right)^{2}, 
\end{equation}
with 
\begin{eqnarray}
N^{2} (r) &=& -8 M G + \frac{r^{2}}{l^{2}} + \frac{16 J^{2}  G ^{2} }{
r ^{2}} \label{n2}\\ 
N_{\phi} (r) &=& - \frac{4 J G}{ r^{2}} \label{nf}.
\end{eqnarray}

The horizon of the black hole is located at $r_{+}$, defined by:
\begin{equation}\label{rpdef}
r_{\pm}^{2} = 4 M G l^{2} \left(1 \pm  \sqrt{1-J^{2}/ (M l)^{2}} \right).
\end{equation}

Moreover, the area, angular momentum and surface gravity on the
horizon are respectively:
\begin{eqnarray}
{\cal A} &=& 2\pi r_{+},\label{area}\\
\Omega &=&  \frac{4 J G}{r_{+}^{2}}, \label{angula}\\
\kappa &=& \frac{r_{+}^{2}- r_{-}^{2} }{l^{2} r_{+} }  \label{surfgra}.
\end{eqnarray}

\bigskip

At spatial infinity ($ r\rightarrow \infty $), the BTZ metric has the 
two sets of asymptotic Killing fields found by Brown
and Henneaux \cite{BHCC}. A
simple basis for these vector fields is:
\begin{eqnarray}
\xi _{n}^{+} &=&  \frac{1}{2} e^{i n (t/l+\phi)} \left(l, - i r n
, 1 \right)
+ \left( O (\frac{1}{r^{2}}),  O(\frac{1}{r}) ,   O (\frac{1}{r^{2}})\right) \label{xiplus}\\
\xi _{n}^{-} &=&   \frac{1}{2} e^{i n (t/l- \phi)} \left( l , - i r n,
- 1 \right) + \left( O (\frac{1}{r^{2}}),  O(\frac{1}{r}) ,   O (\frac{1}{r^{2}})\right).\label{ximinus}
\end{eqnarray}

The overall normalization is chosen such that the Lie bracket between
two vectors gives properly normalized {\it Diff}$(S^{1})$
algebras,
\begin{eqnarray}\label{algdiff1}
i\left\{ \xi _{m}^{+} , \xi _{n}^{+} \right\} &=&  (m-n) \xi _{m+n}^{+}\\
i\left\{ \xi _{m}^{-} , \xi _{n}^{-} \right\} &=&  (m-n) \xi _{m+n}^{-}\\
i\left\{ \xi _{m}^{+} , \xi _{n}^{-} \right\} &=& 0.
\end{eqnarray}

We then use the metric (\ref{BTZ}) and the vector fields
(\ref{xiplus}) and (\ref{ximinus}) in equation (\ref{findth}).  
After performing the integration over $\phi$ from $0$ to $2\pi $ and
taking the limit $r\rightarrow \infty$, we get:
\begin{eqnarray}
\left[Q_{m}^{+},Q_{n}^{+} \right] &=& - i \delta _{m+n}
\left(m (lM-J) + 
m^{3} \frac{l}{8G} \right)\label{dire1}\\
\left[Q_{m}^{-},Q_{n}^{-} \right] &=& - i \delta _{m+n}
\left(m (lM+J) + 
m^{3}  \frac{l}{8G} \right) \label{dire2} \\
\left[Q_{m}^{+},Q_{n}^{-} \right] &=& 0.
\end{eqnarray}

The left hand sides of equations (\ref{ec1}) and (\ref{ec2}) can then
be directly read from equations (\ref{dire1}) and (\ref{dire2})
for $m=-n=1$ and $m=-n=2$. A simple rearrangement then gives:
\begin{eqnarray}
Q_{0}^{+} &=& \frac{1}{2} \left(lM-J \right) + \frac{l}{16 G} \label{resw1}\\
Q_{0}^{-} &=& \frac{1}{2} \left(lM+J \right) + \frac{l}{16 G}
\label{resw2}\\
c^{+} &=& c^{-} = \frac{3 l}{ 2 G} \label{resw3}
\end{eqnarray}
in perfect agreement with \cite{Banados:wn,BHCC}.

The Strominger's derivation \cite{Str} of the black hole entropy using
the Cardy formula is almost straightforward. There is just a subtle
point which concerns the effective values of $Q_{0}^{\pm}$ to be used
in equation (\ref{cardy}). As 
discussed in \cite{Banados:wn}, the adS$_{3}$ metric is recovered from
the BTZ solution 
(\ref{BTZ}) not by taking $M=0$ but  instead
$M=-\frac{1}{8G}$ (and $J=0$). Indeed, the generators (\ref{resw1})
and (\ref{resw2}) came naturally normalized such that
$\left.Q_{0}^{\pm} \right|_{adS_{3}}= 0$. However,
there is no physical
solution (a spacetime
without naked singularities) between $M=-\frac{1}{8G}$ and
$M=0$. Therefore, the continuous 
spectrum of solutions starts with $M\geq 0$ (and  $ J^{2}\leq
 M^{2}l^{2} $). 
Following Strominger, the states responsible for the entropy are then
fluctuations above the vacuum $M=J=0$. The properly
normalized zero 
modes which annihilate this vacuum are simply $\bar{Q}_{0}^{\pm}=
Q_{0}^{\pm}- \frac{c^{\pm }}{24}$. Using these in the Cardy formula
(\ref{cardy}), we then recover the famous result:
\begin{equation}\label{btzcardy}
S = 2 \pi \sqrt{\frac{c^{+}}{6} \bar{Q}_{0}^{+}} + 2
\pi \sqrt{\frac{c^{-}}{6} \bar{Q}_{0}^{-}} =
\frac{{\cal A}}{4G}.
\end{equation}

\bigskip 

Note 
finally that if we work with the bracket
(\ref{nbrack}) instead of the formula (\ref{findth}) to compute the
left-hand side 
of equations (\ref{ec1}) and (\ref{ec2}), we cannot reproduce
the known results (\ref{resw1})-(\ref{btzcardy}). In fact, a direct
calculation gives $\tilde{Q}^{\pm}_{0} = \frac{r^{2}}{4lG}
+ \frac{Ml}{2} + \frac{l}{16G}$ and $\tilde{c}^{\pm}=\frac{3 l}{ 2
G}$. Although 
the central charges are unchanged,  the charges
$\tilde{Q}^{\pm}_{0}$ are equal to each other (the dependence in $J$ is lost)
and diverge at 
$r\rightarrow 
\infty $. Now, even when we use the renormalized
$\bar{Q}^{\pm}_{0}=\frac{Ml}{2}$ in the Cardy formula, we cannot
recover the semi-classical value for the entropy.

\subsection{The BTZ black hole at the horizon}\label{horizonBTZ}

The idea is to repeat the above exercise on the horizon. To
proceed, we first need to identify a family of  {\it Diff}$(S^{1})$
vector fields which preserve the structure of the horizon. Following
the work of Carlip \cite{Carlip:1999cy} summarized in the
appendix B, we consider a set of diffeomorphisms given by
(see equation (\ref{diffeo})):
\begin{equation}\label{familydiff}
\xi _{n}^{\mu } = T_{n} \chi ^{\mu } + R_{n} \rho ^{\mu } 
\end{equation}
where $\chi ^{\mu} = (1,0,\Omega
)$ is the Killing field whose norm vanishes on the horizon. The vector
$\rho^{\mu }$ is defined by formula
(\ref{defrho}). 

Let us then consider the general ansatz (obviously motivated by the
search of a {\it 
Diff}$(S^{1})$ algebra) for the functions $T_{n}$,
\begin{equation}\label{genralan}
T_{n} = \frac{1}{\alpha  +\Omega } e^{in\left(\phi + \alpha  t  + f
(r)\right)}, 
\end{equation}
and therefore (to leading order, see equation (\ref{implies})),
\begin{equation}\label{rn}
R_{n} = -in \frac{\alpha  +\Omega }{\kappa } T_{n}.
\end{equation}

The requirement that the vector fields $\xi _{n}$ satisfy 
the  {\it Diff}$(S^{1})$ algebra (\ref{diff1}) first, fixes the overall
normalization and second, implies that 
the arbitrary function $f (r)$ has to be well-behaved on
the horizon when Schwarzschild coordinates are used. That is $f (r) $
should be {\it finite} as $r\rightarrow r_{+}$.  
This condition is indeed a consequence of equation (\ref{req}). In fact, in 
Schwarzschild coordinates, $\rho ^{\mu}\nabla_{\mu} T_{n}
\sim (r-r_{+}) \partial_{r}  f (r) $ vanishes on the horizon
only if $f (r)$ is well defined on ${\cal H}$.
The constant parameter $\alpha $ will be discussed later.

So, using the vector fields (\ref{familydiff}) in equation
(\ref{findth}), integrating over $\phi$ and taking the limit
$r\rightarrow r_{+}$, we get:
\begin{equation}\label{fhor}
\left[Q_{m} ,Q_{n} \right] = - i \delta _{m+n}
m^{3} \frac{{\cal A}}{8\pi G} \frac{\alpha  +\Omega }{\kappa}.
\end{equation}

We see that the dependence in the arbitrary function $f (r)$,
completely disappears from the final expression (\ref{fhor}). Note
however that the regularity of $f (r)$ on the horizon was required 
in order to obtain this result.
Next, we use this formula (\ref{fhor}) together with equations
(\ref{ec1}) and (\ref{ec2})  to single out the zero-mode and the
central charge:
\begin{eqnarray}
Q_{0} &=& \frac{{\cal A}}{16 \pi G} \frac{\alpha  +\Omega }{\kappa}
\label{simplget1}\\ 
c &=&  \frac{3 {\cal A}}{2 \pi G} \frac{\alpha  +\Omega }{\kappa}.
\label{simplget2}
\end{eqnarray}

Using finally these results in the 
Cardy formula (\ref{cardy}), we obtain:
\begin{equation}\label{cardyh}
S = \frac{{\cal A}}{4 G} \frac{\alpha  +\Omega }{\kappa}.
\end{equation}

This coincides with the usual semiclassical formula only for
\begin{equation}\label{assign}
\alpha =\kappa-\Omega,
\end{equation}
which was imposed by hand in \cite{Carlip:1999cy}.

One can however intuitively argue in favor of this
assignment (\ref{assign}): 
the natural variables on the horizon are $v_{-}=\phi-\Omega t$ (since $\chi
^{\mu } \nabla _{\mu } v_{-}= 0$) and $t$, with periods $2\pi$ and
$\frac{2\pi }{\kappa}$ respectively. 
However, this period $\frac{2\pi }{\kappa}$ is based on the 
semi-classical computation of Gibbons and Hawking \cite{GH}, 
which we would like to avoid in our symmetry-based calculation of
thermodynamical quantities. An independent way for understanding
equation (\ref{assign}) is expected to exist, but unknown to the author.

\bigskip 

In the rest of this 
subsection we compare the results (\ref{simplget1}), (\ref{simplget2})
and (\ref{cardyh})
with previous  
works \cite{Carlip:1998wz,Carlip:1999cy,Dreyer:2001py}. 
First, our equation (\ref{fhor}) is in essence equivalent to equation (3.8) of
\cite{Carlip:1998wz} (with $\alpha +\Omega = \frac{2\pi }{T }$). 
However, the construction of our zero
mode (\ref{simplget1}) differs from the one given in the reference
\cite{Carlip:1998wz} (which 
was later criticized in \cite{Park:1999tj}).

Let us now use the general
ansatz (\ref{genralan}) in the modified bracket
(\ref{nbrack}). After integrating over $\phi$ and taking the limit
$r\rightarrow r_{+}$, we get: 
\begin{equation}\label{nhor}
\left[ \left[ Q_{m} ,Q_{n} \right] \right] = - i\delta _{m+n}  \left( m
\frac{{\cal A}}{4\pi G} \frac{\kappa}{\alpha  +\Omega } + 
m^{3} \frac{{\cal A}}{8\pi G} \frac{\alpha  +\Omega }{\kappa} \right).
\end{equation}
Then using the same argument above equations (\ref{ec1}) and
(\ref{ec2}), we find
\begin{eqnarray}
\tilde{Q}_{0} &=& \frac{{\cal A}}{8\pi G} \frac{\kappa}{\alpha  +\Omega } +
\frac{{\cal A}}{16 \pi G} \frac{\alpha  +\Omega }{\kappa}
\label{nsimplget1}\\ 
\tilde{c} &=&  \frac{3 {\cal A}}{2 \pi G} \frac{\alpha  +\Omega }{\kappa}.
\label{nsimplget2} 
\end{eqnarray}

First, a naive use of $\tilde{Q}_{0}$ and $\tilde{c}$ in the Cardy
formula (\ref{cardy}) does not work. 
Now, even if we assume that equation
(\ref{assign}) holds, the equation (\ref{cardy}) gives $S=\sqrt{3}
\frac{{\cal A}}{4G}$. 

In the references \cite{Carlip:1999cy,Dreyer:2001py}, the zero-mode
was computed 
using the Komar integral (\ref{komar}). The result is indeed
consistent with equation (\ref{nsimplget1}), up to a
constant shift: 
\begin{equation}\label{komeval}
J_{0} = \tilde{Q}_{0} - \frac{\tilde{c}}{24} = \frac{{\cal A}}{8\pi G}
\frac{\kappa}{\alpha  +\Omega }.
\end{equation}

Then, if we use this effective $J_{0}$ in the Cardy formula, we find $S =
\sqrt{2}\frac{{\cal A}}{4 G}$, independently of the value of
$\alpha$. This $\sqrt{2}$-anomaly was first pointed
out in \cite{Dreyer:2001py}. Indeed, up to a change of variable described in appendix C,
the ``extended symmetries'' proposed in \cite{Dreyer:2001py} are nothing
but a special case of equation (\ref{genralan}), namely for $\alpha =
\Omega $. 

Finally, in Carlip second paper \cite{Carlip:1999cy}, the relation $\alpha =
\kappa -\Omega $ is assumed from the beginning and therefore equation 
(\ref{komeval})
 gives $\frac{{\cal A}}{8\pi G}$. In order to avoid the $\sqrt{2}$
anomaly discussed before, another $- \frac{c}{24}$ was then removed
from $J_{0}$. From 
our point of view, where the global normalization of $\tilde{Q}_{0}$ is fixed
by the form of the algebra (\ref{isomorphic}),
this last step looks a little artificial.

Therefore, only in the case where $\frac{c}{24}$ is removed twice from
the ``natural'' generator (\ref{nsimplget1}) and equation
(\ref{assign}) is assumed to hold, the bracket (\ref{nbrack}) succeeded
in reproducing the correct black hole entropy. In that sense, our
derivation based on equation (\ref{findth}) seems more
straightforward since it only depends on the equality  (\ref{assign}).

\subsection{The first law of thermodynamics}\label{flaw}

The first law of black hole thermodynamics is also encoded in
equation (\ref{algebra}). This can indeed be proven in a general way
using covariant symplectic methods 
 \cite{ACK,Francaviglia:2001ww}. Now for a given black
hole, we can check this first law in a simple way. The basic idea is
to consider the metric $g_{\mu \nu }$ as a functional of
$r_{+}$ and $J$ (instead of $M$ and $J$). The BTZ metric is therefore
given by equations (\ref{BTZ}) and (\ref{nf}), together with:
\begin{equation}\label{newn2}
N^{2}=\frac{1}{l^{2}r^{2}} \left(r^{2}-r_{+}^{2} \right)\left(
r^{2}-\left(\frac{4 G l J}{r_{+}} \right)^{2}\right).
\end{equation}
In this case, a variation of $g_{\mu \nu}$ is given by
\begin{equation}\label{dg}
\db g_{\mu \nu } = \frac{\partial g_{\mu \nu }}{\partial r_{+}} \db r_{+} + \frac{\partial g_{\mu \nu }}{\partial J} \db J.
\end{equation}

Then, using the timelike Killing vector
 $\xi^{t}= \left(1,0,0 \right)$ and the explicit form (\ref{dg}) of $\db g_{\mu
\nu }$  in equation (\ref{algebra}), we get
\begin{equation}\label{firlaw}
\db Q_{t} = \left(-\frac{4G J^{2}}{r_{+}^{3}}+ \frac{r_{+}}{4 G l^{2}}
\right) \db r_{+} + \frac{4 J G}{r_{+}^{2}} \db J.
\end{equation}

Using then that\footnote{It is easy to check that $\xi^{t}$ is
associated with the mass of the black hole. In fact, using that $\xi^{t}=
\frac{1}{l} ( 
\xi^{+}_{0}+\xi^{-}_{0} ) $ (see equations (\ref{xiplus}) and
(\ref{ximinus})),  we get  $Q_{t}= \frac{1}{l} (
 Q_{0}^{+} +  Q_{0}^{-} )= M +\frac{1}{8G}$ (see equations
(\ref{resw1}) and 
(\ref{resw2})).} $\delta Q_{t}= \delta
M$ together with the definitions (\ref{area}), (\ref{angula}) and
(\ref{surfgra}), the equation (\ref{firlaw}) can 
be rewritten as:
\begin{equation}\label{fff}
\db M = \frac{1}{8\pi G} \kappa \db A + \Omega \db J.
\end{equation}
Note that this result is independent of the boundary considered in
the equation (\ref{algebra}). In fact, the integral 
does not depend on $r$ (when computed with $\xi^{t}$ and with the BTZ
metric).

\subsection{Four dimensional black holes}\label{fdblack}

Let us consider the Kerr-adS$_{4}$ black hole \cite{Carter:ks} (see
also \cite{Kostelecky:1995ei}-\cite{Awad:1999xx}):
\begin{eqnarray}
ds^{2} &=& -\frac{\Delta _{r}}{\Sigma } \left(dt - \frac{a \sin^{2}
(\theta)}{\Xi}
d\phi \right)^{2} + \frac{\Sigma }{\Delta _{r}} dr^{2} + \frac{\Sigma
}{\Delta _{\theta }}d \theta ^{2} \nonumber\\
 & & + \frac{\Delta _{\theta }\sin^{2}
(\theta) }{\Sigma } \left(a dt -\frac{r^{2}+a^{2}}{\Xi} d\phi
\right)^{2}\label{ads4} 
\end{eqnarray}
with,
\begin{eqnarray}
\Delta _{r} &=& \left(r^{2}+a^{2} \right)\left(1+\frac{r^{2}}{l^{2}}
\right) - 2 M G r \nonumber\\
&=& \Delta '_{r_{+}} (r-r_{+}) + \left(6
\frac{r_{+}^{2}}{l^{2}}+1+\frac{a^{2}}{l^{2}} \right) (r-r_{+})^{2}
\nonumber \\
& & + 4 \frac{r_{+}}{l^{2}} (r-r_{+})^{3} + \frac{1}{l^{2}} (r-r_{+})^{4}\\
\Delta '_{r_{+}} &=& \left.\frac{\partial \Delta _{r} }{ \partial r}
\right|_{r=r_{+}} = r_{+} \left(3 \frac{r_{+}^{2}}{l^{2}} +
\frac{a^{2}}{l^{2}} + 1 - \frac{a^{2}}{r_{+}^{2}} \right)\\
\Delta _{\theta } &=& 1 -\frac{a^{2}}{l^{2}} \cos^{2} (\theta )\\
\Sigma &=& r ^{2} + a^{2} \cos^{2} (\theta )\\
\Xi &=& 1 -\frac{a ^{2}}{l^{2}},
\end{eqnarray}
where $r_{+}$ is the highest root of $\Delta _{r}$.

The area, angular velocity and the surface gravity (equation
(\ref{limi})) of the horizon are \cite{Wald:rg}:
\begin{eqnarray}
{\cal A} &=& \int_{r=r_{+}} \sqrt{g_{\theta \theta }g_{\phi \phi }}\ \ 
d\theta d\phi = 4 \pi \frac{r_{+}^{2}+a^{2}}{\Xi} \label{a4}\\
\Omega &=&-\left.\frac{g_{t\phi }}{g_{\phi \phi }} \right|_{r=r_{+}} =
\frac{a \ \Xi}{r_{+}^{2}+a^{2}}\label{om4}\\
\kappa &=& \frac{\Delta '_{r_{+}}}{2 (r_{+}^{2}+a^{2})} \label{kappa4}
\end{eqnarray}

The mass and angular momentum are given by\footnote{We used the
superpotential of Katz, Bi\u{c}{\'a}k and Lynden-Bell \cite{KBL} (see also
\cite{JS2}) to compute ${\cal M}$ and ${\cal J}$ (respectively associated
with the Killing vectors $\partial_{t}$ and $-\partial_{\phi}$). The background
metric to be used is the adS$_{4}$ spacetime seen by a rotating observer
which is obtain from (\ref{ads4}) by setting $M=0$ \cite{Caldarelli:1999xj}.}:
\begin{eqnarray}
{\cal M} &=& \frac{M}{\Xi}\label{m4}\\
{\cal J } &=& \frac{a M}{\Xi^{2}}. \label{j4}
\end{eqnarray}

Note that the Kerr solution, together with its charges and its thermodynamical
quantities can be obtained from the metric (\ref{ads4}) by
taking the limit $l\rightarrow \infty$. The Schwarzschild solution
then follows after setting $a$ to zero. Moreover, all our following
results will remain valid in these limits.

Following section \ref{horizonBTZ}, the asymptotic diffeomorphisms on
the horizon are given by equation (\ref{familydiff}).
The null Killing vector field is now $\chi^{\mu } = (1, 0, 0,
\Omega )$. We choose a natural four-dimensional extension of the ansatz
(\ref{genralan}) for the functions $T_{n}$ 
\begin{equation}\label{tn4}
T_{n} = \frac{1}{\alpha  +\Omega } e^{in\left(\phi + \alpha  t  + f
(r,\theta)\right)}.
\end{equation}

The normalization of $T_{n}$ has been fixed by the algebra
(\ref{diff1}). Moreover, we can verify that the constraint
(\ref{req}) again requires that the arbitrary function $f (r,\theta )$
has to be 
well-defined (finite) in the limit $r\rightarrow r_{+}$.

We next use the metric (\ref{ads4}) and the vector fields
$\xi_{n}^{\mu }$ (derived from equations (\ref{familydiff})
and (\ref{tn4})) in the formula (\ref{findth}).
After a tedious calculation,
which involves integrating over $\phi$ (from $0$ to $2\pi$) and over 
$\theta$ (from  $0$ to $\pi$) and taking 
the limit $r\rightarrow r_{+}$, we finally get:
\begin{equation}\label{same}
\left[Q_{m} ,Q_{n} \right] = - i \delta _{m+n}
m^{3} \frac{{\cal A}}{8\pi G} \frac{\alpha  +\Omega }{\kappa},
\end{equation}
which is identical to (\ref{fhor}). We can then follow the same
argument to conclude that the semiclassical entropy is again recovered
if equation (\ref{assign}) is satisfied. 

Note that this derivation of the Kerr-adS$_{4}$ black hole entropy remains
equally valid for the Kerr and the Schwarzschild solutions since the
limits $l\rightarrow \infty $ and $a\rightarrow 0$ are always well
defined. 

We finally check that equation
(\ref{algebra}) indeed reproduces the first law of thermodynamics. 
To proceed, we use the same trick outlined by equation (\ref{dg}),
that is, we re-express the metric (\ref{ads4}) as a function of $r_{+}$
and $a$ in order to compute $\db g_{\mu \nu}$. Using this
varied metric and the timelike Killing vector $\xi ^{t}:= (1,0,0,0)$
in equation (\ref{algebra}), we get:
\begin{eqnarray}
\db Q_{t} &=& \frac{\Delta '_{r_{+}}}{2 r_{+} \Xi G}  \db r_{+} +
\frac{a \left(1+\frac{r_{+}^{2}}{l^{2}} \right) \left(3
\frac{r_{+}^{2}}{l^{2}} + \frac{a^{2}}{l^{2}} +2 \right)}{2 r_{+}
\Xi^{2} G} \db a \nonumber \\
&=& \frac{1}{8\pi G} \kappa \db A + \Omega \db J.\label{dq4}
\end{eqnarray}
We used equations (\ref{a4}), (\ref{om4}), (\ref{kappa4}) and
(\ref{j4}) in the second line. Note moreover that the computed integral
(\ref{algebra}) is independent of $r$, and therefore remains valid on any
boundary (at spatial infinity or on the horizon).

It is important to compare the result (\ref{dq4}) with the first law
derived explicitely in the reference \cite{Caldarelli:1999xj}. We first
need to rewrite $\db Q_{t}$ in term of the mass (\ref{m4}). The
simplest way to
proceed is to 
re-consider the metric as a functional of $M$ and
$a$. Using now $\db g = \frac{\partial g}{\partial M}  \db M +
\frac{\partial g}{\partial a}  \db a $ in equation (\ref{algebra}), we find:
\begin{equation}\label{dqqqq}
\db Q_{t} = \db {\cal M} + \frac{a {\cal M}}{l^{2} \Xi} \db a
\end{equation}

Combining equations (\ref{dq4}) and (\ref{dqqqq}), we find after some
simple algebra:
\begin{equation}\label{fid}
\db \left( \frac{{\cal M}}{ \Xi}  \right) = \frac{1}{8\pi G} \kappa
\db A + \tilde{\Omega} \db J,
\end{equation}
where  $\tilde{\Omega} = \Omega +\frac{a}{l^{2}}$ is the angular
velocity at infinity \cite{Hawking:1998kw}. We then recovered the
result given in \cite{Caldarelli:1999xj}.

\section{Conclusion}\label{conc}

We have presented a new method to derive the zero-mode $Q_{0}$ and the
central charge $c$ of a given black hole solution. The entropy
computed using the Cardy formula then coincides with the
Bekenstein-Hawking formula. A simple derivation of the first law of
black hole thermodynamics using our framework was also given.

We found a one-parameter family of {\it Diff}$(S^{1})$ algebras which
preserve the Carlip boundary conditions on the horizon. However, only one
of them gives the correct entropy, namely for $\alpha =\kappa -\Omega $.
We expect that there exists an additional natural constraint to be imposed on
the horizon which would single out this particular {\it Diff}$(S^{1})$
algebra. Finally, if new {\it Diff}$(S^{1})$ algebras are found on the
horizon, our formula (\ref{findth}) could then be used to check their
ability to derive the black hole entropy.

We have then shown that the Cardy formula is
able to handle a microscopical black hole entropy calculation with
success. We hope 
to better understand in the future the quantum conformal theory on the
horizon which is responsible for this remarkable result.

\bigskip

{\bf Acknowledgments.}

I benefited from enlightening correspondence with G. Barnich, F. Brandt
and S. Carlip and from discussions with G. Arutyunov and E. Lopez.

\section*{Appendix A: Some comments on equation (\ref{findth})}
\renewcommand{\theequation}{A.\arabic{equation}}
\setcounter{equation}{0}

The purpose of this first appendix is to compare the equation 
(\ref{findth}) with similar results given in the literature. 

 Let us first assume that we have a natural background metric denoted
by $\overline{g}_{\mu \nu }$ on a given boundary 
(for instance flat metric at spatial infinity). 
We can then normalize the charges $Q_{n}$ such that $Q_{n}[\overline{g}] = 0$.
As a consequence of the formula (\ref{isomorphic}), the equation
(\ref{findth}) evaluated on this 
background $\overline{g}_{\mu \nu }$ gives then the central
charge. We can then compare this result with other formulas derived in
the literature.

First, it is straightforward to check that on-shell (${\cal R}_{\mu
\nu } = \frac{2}{D-2}\Lambda g_{\mu \nu }$), equation (\ref{findth})
evaluated on $\overline{g}_{\mu \nu }$ perfectly agrees with the
central charge formula found by Koga \cite{Koga:2001vq} using the covariant
symplectic methods. This is not really a surprise since the
``covariant Regge-Teitelboim'' method \cite{Si} used to derive equation
(\ref{algebra}) is equivalent to the symplectic techniques developed by
Ashtekar \cite{ABR}, Wald \cite{Wa} and collaborators. This was shown
in \cite{Barnich:2001jy}, and will be 
detailed also in \cite{JS4}. 

A similar formula for the central charge was derived by Barnich and
Brandt \cite{Barnich:2001jy}.
There is again a good agreement with Koga's and our
results, up to the term $(\overline{\nabla} ^{\rho} \xi
^{\nu }_{m} + 
\overline{\nabla} ^{\nu} \xi ^{\rho }_{m}) (\overline{\nabla} ^{\mu}
\xi_{ \rho n} + \overline{\nabla}_{\rho } \xi ^{\mu }_{n} ) dS_{\mu
\nu }$. 
Although this extra contribution vanishes for all known
examples\footnote{G. Barnich and F. Brandt, {\it private
communication}.}, it would 
be of interest to understand its meaning.

An alternative to equation (\ref{findth}) for the bracket between two
charges was derived by Carlip in \cite{Carlip:1999cy} (see equation
(3.3) (and (3.6)) of that paper). This bracket, also used by Jing and 
Yan \cite{JY} and by  Dreyer,
Ghosh and Wi\'{s}niewski \cite{Dreyer:2001py}, is denoted here by
$[[\cdot,\cdot ]]$ : 
\begin{eqnarray}
\left[ \left[ Q_{m}, Q_{n} \right] \right] = \int_{\it bd}
\frac{\sqrt{\left|  {g} \right|}}{16 \pi G} 
 & & \left[\xi_{m} ^{\mu } {\nabla}_{\rho } \left( \nabla ^{\rho } \xi_{n}
^{\nu } - \nabla ^{\nu } \xi_{n}^{\rho } \right)-
\xi_{n} ^{\mu } {\nabla}_{\rho } \left( \nabla ^{\rho } \xi_{m}
^{\nu } - \nabla ^{\nu } \xi_{m}^{\rho } \right) \right. \nonumber\\
& & + \left. \xi _{m}^{\mu} \xi _{n}^{\nu} {\cal L} 
- (\mu \leftrightarrow \nu) \right] dS_{\mu \nu }\label{nbrack}
\end{eqnarray}

On-shell, the Lagrangian is related to the Ricci tensor by ${\cal R}_{\mu
\nu } =\frac{1}{2}g_{\mu \nu } {\cal L}$. 
Then, it can be shown 
that the difference between equation (\ref{findth}) and equation
(\ref{nbrack}) is  (up to total derivatives analogous to (\ref{totder}))
\begin{equation}\label{ndiff}
\int_{\it bd}\frac{\sqrt{\left|  {g} \right|}}{16 \pi G}
 \left[ \nabla ^{\mu } \left( \xi_{m}^{\rho }  {\nabla}_{\rho } \xi
_{n}^{\nu}  -   \xi_{n}^{\rho }  {\nabla}_{\rho } \xi
_{m}^{\nu}\right)
- (\mu \leftrightarrow \nu) \right] dS_{\mu \nu }.
\end{equation}

This is nothing but the (1/2) Komar superpotential \cite{Wald:rg}
(also called Noether charge ${\mathbf{Q}} [ g ,\xi]$ in \cite{Wa})
\begin{equation}\label{komar}
J[\xi] = \int_{\it bd}\frac{\sqrt{\left|  {g} \right|}}{16 \pi G}
 \left[ \nabla ^{\mu } \xi^{\nu } - (\mu \leftrightarrow \nu)
\right] dS_{\mu \nu }, 
\end{equation}
evaluated for the Lie bracket of the two
vector fields $\xi _{m}^{\mu}$ and $\xi _{n}^{\nu} $. 
As pointed out in a footnote by Koga \cite{Koga:2001vq},
the mismatch comes from the
equation (3.2)  of the reference \cite{Carlip:1999cy}. Indeed, if we
assume that 
$\delta _{\xi}$ acts on the metric but {\it not} on the parameters,
this equation (3.2) has to be modified to 
$\delta _{\xi_{m}} J[\xi _{n}] = (\xi_{m}\cdot \ex + \ex \xi_{m}\cdot
) J[\xi _{n}] - J[\{\xi_{m}, \xi _{n} \} ]$ (see for instance equation
(4.8) of \cite{Si} for the analogous example of Yang-Mills
theory). The last 
term  $ J[\{\xi_{m}, \xi _{n} \} ]$  gives  the additional
contribution (\ref{ndiff}) on the horizon. In the examples of section
\ref{exa}, the extra term (\ref{ndiff}) does
modify the zero-mode $Q_{0}$ but not the central charge.

\section*{Appendix B: Local Killing horizons : the Carlip's approach}
\renewcommand{\theequation}{B.\arabic{equation}}
\setcounter{equation}{0}

The goal of this appendix is to summarize part 
of the Carlip's results  \cite{Carlip:1999cy}
on the asymptotic symmetries of (local) Killing horizons. 
Although we present this work in a slightly rephrased way, the
important results  
remain unchanged. We refer to the original paper for a
detailed discussion. 

Let us assume that a locally isolated Killing horizon can be
characterized by some vector field $\chi^{\mu}$, whose norm 
\begin{equation}\label{bhhorizon}
\chi^{2} = 0.
\end{equation}
vanishes on the horizon.
We work on the stretched horizon, and we will take the limit $\chi^{2} =
0$ at the end of the calculations. 

In the following, we just need to
assume that the vector field $\chi^{\mu}$ is a Killing vector on
the horizon and a conformal Killing vector to first order\footnote{This is little weaker than the original
assumption of Carlip that the vector $\chi^{\mu}$ is a Killing vector
in a neighborhood of the horizon.}. Concretely, it is enough to
require that: 
\begin{equation}\label{weak}
\nabla_{( \mu } \chi_{\nu ) } = O (\chi ^{2}) g_{\mu \nu } + O (\chi ^{4}).
\end{equation}

Following Carlip's paper, the vector normal to the horizon is defined by
\begin{equation}\label{defrho}
\rho _{\mu } := - \frac{1}{2\kappa } \nabla_{\mu }  \chi ^{2}.
\end{equation}
Then, the norm of $\rho _{\mu }$ vanishes on the horizon
(null hypersurface)
\begin{equation}\label{normzero}
\rho ^{2} = O (\chi^{2}).
\end{equation}

Now, the normalization factor $\kappa$ in the definition of $\rho
_{\mu }$ (\ref{defrho}) is
fixed by the requirement that close to the horizon,
\begin{equation}\label{normrho}
\rho ^{2} = -\chi^{2} + O (\chi^{4}).
\end{equation}
The overall normalization of $\chi$ is therefore the only
remaining free parameter.  
Moreover, from equations (\ref{weak}) and (\ref{defrho}), $\kappa$ can be
identified with the surface gravity associated with $\chi$, 
\begin{equation}\label{limi}
\kappa ^{2} := -\lim_{\chi^{2} \rightarrow 0} \left( \frac{\nabla_{\sigma
}\chi^{2} 
\nabla^{\sigma  }\chi^{2}}{4 \chi^{2}}\right) = -\lim_{\chi^{2} \rightarrow 0} \left(
\frac{ a^{\sigma } a_{\sigma }}{ \chi^{2}}\right),
\end{equation}
with the acceleration given by $a^{\sigma} := \chi^{\mu}\nabla_{\mu
}\chi^{\sigma}$. 

We can use equations  (\ref{weak})  and (\ref{defrho}) to derive
some useful identities, 
\begin{eqnarray}
\rho ^{\mu } \chi_{\mu } &=& O (\chi ^{4}),\label{rxz1}\\ 
{\cal L}_{\chi} \rho_{\mu } &=& \nabla_{\mu } (\rho \cdot  \chi) = O
(\chi ^{2}) \rho_{\mu }  \label{rxz2}\\
{\cal L}_{\chi} \rho^{\mu } &=& O (\chi ^{2})\rho^{\mu }.\label{rxz3}
\end{eqnarray}

The next step is to identify a set of  diffeomorphisms of the metric which
preserve some of the above structure. In particular and following
Carlip, we
require that the position and the normal direction of the black hole
(equations (\ref{bhhorizon}) and (\ref{defrho})) remain unchanged
under $g_{\mu \nu} \rightarrow g_{\mu \nu} + {\cal L}_{\xi }g_{\mu
\nu}$, that is: 
\begin{eqnarray}
\delta_{\xi} \chi ^{2} &=&  \chi ^{\mu }\chi ^{\nu }{\cal L}_{\xi
} g_{\mu \nu} = O (\chi ^{4}) \label{prese1} \\
\delta _{\xi} \rho _{\mu} &=& - \frac{1}{2\kappa} \nabla _{\mu }
\left( \chi ^{\nu}\chi ^{\sigma }{\cal
L}_{\xi } g_{\nu \sigma}  \right) = \rho _{\mu} O (\chi ^{2})\label{prese2}
\end{eqnarray}

For the ansatz 
\begin{equation}\label{diffeo}
\xi ^{\mu } = R \rho ^{\mu } + T\chi ^{\mu },
\end{equation}
the equations (\ref{prese1}) and  (\ref{prese2}) are satisfy if 
\begin{equation}\label{implies}
{\cal L}_{\chi} T + \kappa R = O (\chi ^{2}).
\end{equation}
This equation then gives $R$ in term of $T$.

Finally, the closure of the Lie bracket between two asymptotic parameters
(\ref{diffeo}) requires an additional constraint 
\begin{equation}\label{req}
\rho ^{\mu } \nabla _{\mu } T = O (\chi ^{2}) T.
\end{equation}

Note that the vector field $\xi^{\mu }$ does not need to be well
defined on the horizon itself (this is indeed a coordinate
dependent-statement). However, the associated charges 
should be finite.

\section*{Appendix C: The BTZ metric in Eddington-Finkelstein
coordinates}
\renewcommand{\theequation}{C.\arabic{equation}}
\setcounter{equation}{0}

In Eddington-Finkelstein coordinates, the BTZ black hole is
\cite{Dreyer:2001py}: 
\begin{equation}\label{BTZEF}
ds^{2} = - N^{2}dv^{2}+ 2 dv dr + r^{2} (N_{\varphi }dt + d\varphi )^{2}
\end{equation}
with $N^{2}$ and $N_{\varphi}$ given as before by equations (\ref{n2}) and
(\ref{nf}). 

The change between the Eddington-Finkelstein coordinates $(v,r,\varphi
)$ and the usual Schwarzschild coordinates $(t,r,\phi )$ is simply:
\begin{eqnarray}
v &=& t + A (r)\\
r &=& r\\
\varphi &=& \phi + B (r),
\end{eqnarray}
where the functions $A (r)$ and $B (r)$ satisfy :
\begin{eqnarray}
\partial_{r} A  &=& N^{-2} \label{deq1}\\
\partial_{r} B  &=& - N^{-2} N_{\phi}. \label{deq2}
\end{eqnarray}

Integrating equations (\ref{deq1}) and (\ref{deq2}) explicitely, we find:
\begin{eqnarray}
A &=& \frac{1}{2 \kappa }\left(\log \left(\frac{r-r_{+}}{r+r_{+}}
\right) -\frac{r_{-}}{r_{+}} \log \left(\frac{r-r_{-}}{r+r_{-}}
\right) \right)\\
B &=& \frac{\Omega }{2 \kappa } \left(\log \left(\frac{r-r_{+}}{r+r_{+}}
\right) -\frac{r_{+}}{r_{-}} \log \left(\frac{r-r_{-}}{r+r_{-}}
\right) \right).
\end{eqnarray}

We can then use 
the asymptotic behavior of $A$ and $B$ close to the horizon,
\begin{eqnarray}
A &\stackrel{r\rightarrow r_{+}}{\longrightarrow}& 
\frac{1}{2\kappa } \log (r-r_{+}) + O ((r-r_{+})^{0})\label{Aasy}\\
B &\stackrel{r\rightarrow r_{+}}{\longrightarrow}&
\frac{\Omega }{2 \kappa } \log (r-r_{+}) + O ((r-r_{+})^{0}) \label{Basy}
\end{eqnarray}
to check that the ``extended symmetries'' proposed in
\cite{Dreyer:2001py}\footnote{Our conventions differs by a minus signs
from those of 
\cite{Dreyer:2001py}, for the angular velocity and for the  {\it
Diff}$(S^{1})$ algebra (\ref{diff1}).},
\begin{equation}\label{ddtn}
T_{n} = \frac{1}{2 \Omega } e^{in \left( \varphi + \Omega v -
\frac{\Omega }{\kappa } \log (r-r_{+}))\right)}
\end{equation}
is in the coordinates $t$ and $\phi$, a special case of the general
ansatz (\ref{genralan}) (namely for 
$\alpha =\Omega$).  
We already saw in section \ref{horizonBTZ} that the regular part of
the functions $A (r)$ and $B (r)$
does contribute neither to the zero-mode nor to the central charge and
therefore can be dropped out.

\end{document}